\documentclass{article}
\usepackage{spconfa4,amsmath,graphicx, amsfonts, amssymb, subfigure, xcolor,algorithm,algpseudocode,epstopdf,multirow,threeparttable,multicol,balance, mathtools,cite} 
\DeclareMathSymbol{\shortminus}{\mathbin}{AMSa}{"39}  
\title{Source Localization by Multidimensional Steered Response Power Mapping with Sparse Bayesian Learning}
\name{\fontsize{11}{13}\selectfont Wei-Ting Lai, Lachlan Birnie, Xingyu Chen, Amy Bastine, Thushara D. Abhayapala, Prasanga N. Samarasinghe}
\address{\fontsize{11}{13}\selectfont\text{Audio \& Acoustic Signal Processing Group, The Australian National University, Canberra, Australia}}
\begin{document}
%
\maketitle
\begin{abstract}
We propose an advance Steered Response Power (SRP) method for localizing multiple sources. While conventional SRP performs well in adverse conditions, it remains to struggle in scenarios with closely neighboring sources, resulting in ambiguous SRP maps. We address this issue by applying sparsity optimization in SRP to obtain high-resolution maps.
Our approach represents SRP maps as multidimensional matrices to preserve time-frequency information and further improve performance in unfavorable conditions.
We use multi-dictionary Sparse Bayesian Learning to localize sources without needing prior knowledge of their quantity.
We validate our method through practical experiments with a 16-channel planar microphone array and compare against three other SRP and sparsity-based methods.
Our multidimensional SRP approach outperforms conventional SRP and the current state-of-the-art sparse SRP methods for localizing closely spaced sources in a reverberant room.
\end{abstract}
\begin{keywords}
Source Localization, Steered Response Power, Sparse Representation, Sparse Bayesian Learning
\end{keywords}
\section{Introduction}
\label{sec:intro}

Acoustic source localization in reverberant environments is an active problem in microphone signal processing. Several methods have been developed for source localization, such as subspace-based approaches like MUltiple SIgnal Classification (MUSIC) \cite{music_1,music_2} and ESPRIT \cite{esprit}, time differences of arrival (TDOA) approaches like Generalized Cross-Correlation Phase Transform (GCC-PHAT) \cite{gcc_1,gcc_2} and steered response power (SRP) \cite{gcc_1,srp}, sparsity-based approaches like Orthogonal Matching Pursuit (OMP) \cite{omp}, Sparse Bayesian Learning (SBL) \cite{sbl_1,sbl_2}, as well as learning-based approaches \cite{learning_1,learning_2}.

The SRP method estimates sound source positions by the summation of cross-correlation of all possible microphone pairs. Recently, several approaches have focused on hierarchical search \cite{srpsearch_1,srpsearch_2} and real-time SRP \cite{realtime} to reduce errors and complexity. However, SRP still struggles with localization in scenarios where multiple sources closely spaced in reverberant environments due to an ambiguous SRP map. Some research has overcome this challenge by iterative grid decomposition \cite{zoomin} or sparse fitting \cite{srp_sparse_1,srp_sparse_2}. However, the former approach still requires careful selection for the grid resolution, especially in scenarios with numerous sources. Whereas, the latter performs well in low reverberation scenarios but degrades dramatically as reverberation increases.

In this paper, we propose a method to improve localization performance by following the sparse fitting approach. To reduce high localization errors caused by reverberant environments, we represent the obtained SRP maps as multidimensional matrices to preserve more time-frequency information. We apply multi-dictionary SBL (M-SBL) \cite{sbl_3} as the sparsity optimization method, allowing the proposed method to operate without prior knowledge of the number of sources, thereby adapting to adverse real-world conditions. 
We utilize two sets of candidate grids with different resolutions for SRP mapping and sparsity fitting to further improve efficiency. We validate the proposed method through practical experiments with a planar microphone array, demonstrating improvements in localizing multiple closely spaced sources in a reverberant room compared against other conventional and state-of-the-art methods. Results show our method enhances robustness, allowing for stable performance even when localizing on short duration recordings.

\section{Problem Formulation}
\label{sec:format}

Consider $J$ sound sources, $j {\,=\,} \{1,{\cdots},J\}$, located at positions $\mathbf{y}_j$ that each emit signals $s_j$.
Additionally, consider a microphone array with $M$ elements each positioned at $\mathbf{x}_m$ for $m {\,=\,} \{1,{\cdots},M\}$.
Received signals can be expressed as
\begin{equation}
    p_m(\mathfrak{t})=s_j(\mathfrak{t})*g_m(\mathfrak{t}, \mathbf{y}_j)+v_m(\mathfrak{t}),
    \label{eq:signal1}
\end{equation}
where $p_m(\mathfrak{t})$ denotes the $m^{\text{th}}$-microphone's signal at time $\mathfrak{t}$, $g_m(\mathfrak{t}, \mathbf{y}_j)$ denotes the impulse response from the $j^{\text{th}}$ source to the $m^{\text{th}}$ microphone, and $v_m$ denotes a noise term.

The TDOA between the pair of microphones $(m, m')$ due to the $j^{\text{th}}$ sound source is given by \cite{gcc_1}:
\begin{equation}
    \tau_{m, m^{\prime}}(\mathbf{y}_{j}) 
    = 
    \frac{1}{c} 
    \begin{cases}
        \left\|\mathbf{y}_{j}{-}\mathbf{x}_{m}\right\| - \left\|\mathbf{y}_{j}{-}\mathbf{x}_{m^{\prime}}\right\| & \text {for NF,}\\
        (\mathbf{x}_{m}{-}\mathbf{x}_{m^{\prime}}) \cdot \vec{\mathbf{y}}_{j} & \text {for FF,}
    \end{cases}
    \label{eq: tdoa}
\end{equation}
where $c$ is the speed of sound, NF and FF denote near-field and far-field sound propagation, respectively.

The goal of TDOA-based sound source localization is to find the source locations of $\mathbf{y}_{j}$ from estimated TDOAs \eqref{eq: tdoa} of a microphone array.







\section{Source Localization by SRP} 
\label{sec:theory}

Two common TDOA-based localization methods are GCC-PHAT \cite{gcc_1,gcc_2} and SRP-PHAT \cite{srp}. While GCC-PHAT maximizes the GCC function of a microphone pair $(m, m')$ to estimate TDOAs, SRP-PHAT maximizes the GCC functions of all $L{\,=\,}\binom M2$ microphone pairs denoted by $\ell = \{1,{\cdots},L\}$.

The GCC function at a time-frequency frame is defined as
\begin{equation}
    R_{m, m^\prime}(\tau,k,t) = \Psi_{m{,} m^{\prime}}(k{,}t) P_m(k{,}t) P_{m^{\prime}}^*(k{,}t) e^{j 2\pi k\tau},
    \label{eq:gcc}
\end{equation}
where $k = \{1,{\cdots},K\}$ and $t = \{1,{\cdots},T\}$ index the wave number and the time frame in time-frequency domain, respectively, $K$ and $T$ denote the totals, $P_m(k,t)$ is the short-time Fourier transform of $p_m(\mathfrak{t})$, $(\cdot)^*$ is the complex conjugate, $\Psi_{m, m^{\prime}}(k,t)= 1/|P_m(k,t) P_{m^{\prime}}^*(k,t)|$ is the frequency-dependent weighting function for GCC-PHAT.

Assuming all sound sources belong to a set of $N$ candidate positions $\mathbf{y}_n\in\{\mathbf{y}_1, {\cdots},\mathbf{y}_N\}$, where $N{\,>\,}J$, the SRP function is given as
\begin{equation}
    z_{k,t}(\mathbf{y}_n)=\Re\left({\textstyle\sum_{\ell=1}^L} R_\ell(\tau,k,t)\right),
\label{eq:srp}
\end{equation}
where $\ell\equiv(m,m')$. 

The SRP map is composed of the SRP functions \eqref{eq:srp} of all $N$ candidate positions. When the obtained SRP reaches a local maximum, the candidate position will be an estimated source location. In practice, the source positions $\hat{\mathbf{y}}_j$ are estimated by averaging SRP over time and frequency as follows:
\begin{equation}
    \hat{\mathbf{y}}_j=\arg \max _{n \in N}{z(\mathbf{y}_n)}=\arg \max _{n \in N}{\textstyle\sum_{k=1}^{K}\sum_{t=1}^T} {z_{k,t}(\mathbf{y}_n)},
    \label{eq:srpl10n}
\end{equation}
where $z(\mathbf{y}_n)$ denotes the averaged SRP function of the $n^{\text{th}}$ candidate position.

The SRP method for localization is robust in adverse environments after time-frequency averaging. 
However, multiple sources located close to each other can still lead to an ambiguous SRP map and poor source localization. In the next section we address this issue of closely spaced sources by employing a sparsity-based SRP method with M-SBL \cite{sbl_3}.

\section{Proposed Method}
\label{sec:proposed}
We propose optimizing SRP mapping through the use of M-SBL. We aim for a sparse solution to enforce concentrated peaks in a high resolution SRP map, which allows for distinction between multiple sources placed close to each other.

We define another set of $Q$ candidate positions $\mathbf{y}_q\in\{\mathbf{y}_1, {\cdots},\mathbf{y}_Q\}$, the microphone signal model from $\mathbf{y}_q$ to $\mathbf{x}_m$ can be expressed as a linear regression model by STFT as follows:
\begin{equation}
P_m(k,t)= G_{q,m}(k)S_q(k,t)+\mathcal{N}_m(k,t)
\label{eq:signal3}
\end{equation}
where $P_m(k,t)$ denotes the $m^\text{th}$ microphone signal at the $k^\text{th}$ wave number and the $t^\text{th}$ time frame, $G_{q,m}$ denotes the dictionary matrix, $S_q(k,t)$ denotes the source weight at $\mathbf{y}_q$, and $\mathcal{N}_m(k,t)$ denotes the noise term. 

We assume the reverberation of the environment is unknown, hence the dictionary function $G_{q,m}(k)$ is modeled as Green's functions in case of NF or FF scenarios, respectively, as follows:
\begin{equation}
G_{q,m}(k)= \begin{cases}e^{j k\|\mathbf{x}_m-\mathbf{y}_q\|}/({4 \pi\|\mathbf{x}_m-\mathbf{y}_{q}\|}) & \text { for NF, } \\
e^{-j k \vec{\mathbf{y}}_q \cdot \mathbf{x}_m} & \text { for FF. }\end{cases}
\label{eq:green}
\end{equation}

By combining between equation \eqref{eq:gcc} and equation \eqref{eq:signal3}, we can obtain the corresponding GCC from this candidate position $\mathbf{y}_n$ as follows:
\begin{equation}
    \begin{aligned}
        R_{m,m^{\prime}}&(\tau{,}k{,}t)
        \approx 
        \\
        &\frac{S_q(k{,}t)G_{q{,}m}(k)(S_q(k{,}t)G_{q{,}m^{\prime}}(k))^*}{|S_q(k{,}t)G_{q{,}m}(k)(S_q(k{,}t)G_{q{,}m^{\prime}}(k))^*|}
        e^{j 2\pi k\tau},
        \label{eq:newgcc}
    \end{aligned}
    %
\end{equation}
where the TDOA $\tau$ corresponds one-to-one with the first set of $N$ candidate positions. It is worth noting to clarify the two sets of candidate grids in the proposed method: $N$ is for SRP mapping while $Q$ is for the subsequent sparsity fitting. Following equation \eqref{eq:green}, the TDOA between the microphone pair $(m, m')$ for the candidate position $\mathbf{y}_{n}$, denoted as $\tau_{m, m^{\prime}}(\mathbf{y}_{n})$, can be expressed as follows:
\begin{equation}
    \tau_{m, m^{\prime}}(\mathbf{y}_{n})
    =
    \frac{1}{j 2\pi k}
    \ln\big(G_{n,m^{\prime}}(k)G^*_{n,m}(k)\big).
    \label{eq:newtdoa}
\end{equation}

By combining equation \eqref{eq:srp}, \eqref{eq:newgcc}, and \eqref{eq:newtdoa}, the SRP can be reformulated as:
\begin{equation}
    z_{k,t}(\mathbf{y}_n,\mathbf{y}_{q}) 
    \approx
    \Re\left(\sum_{\ell=1}^LH^*_{n,\ell}(k)\frac{H_{q,\ell}(k)}{|H_{q,\ell}(k)|}\right),
    \label{eq:newsrp}
\end{equation}
where  $H_{n,\ell}(k)=G_{n,m}(k)G^*_{n,m^{\prime}}(k)$ is the relative transfer function of the microphone pair $\ell$ \cite{retf}.

We define vectors $\mathbf{a}_\ell(k){\,\in\,}\mathbb{C}^{N {\times} 1}$ and $\mathbf{b}_\ell(k){\,\in\,}\mathbb{C}^{1 {\times} Q}$ to represent the right part of equation \eqref{eq:newsrp} for brevity, where
\begin{equation}
\begin{split}
    & \mathbf{a}_\ell(k)
    {=}
    \begin{array}{c}\left[H_{1,\ell}^*(k),{\cdots},H_{N,\ell}^*(k)\right]^{\top},\end{array}
    \\
    & \mathbf{b}_\ell(k)
    {=}
    \begin{array}{c}\left[\frac{H_{1,\ell}(k)}{|H_{1,\ell}(k)|},{\cdots},\frac{H_{Q,\ell}(k)}{|H_{Q,\ell}(k)|}\right].\end{array}
\end{split}
\end{equation}

Then, we can regard the SRP map of all candidate positions as the summation of equation \eqref{eq:newsrp} as follows:
\begin{equation}
    \mathbf{z}_{k,t}
    =
    \Re\left(\mathbf{A}_k\mathbf{B}_k\mathbf{S}_{k,t}\right) +  \boldsymbol{\mathcal{N}}_{k,t},
    \label{eq:newsrp2}
\end{equation}
where $\mathbf{z}_{k{,}t}{\,=\,}[ z_{k,t}(\mathbf{y}_1),{\cdots},z_{k,t}(\mathbf{y}_N)]^\top{\in\,}\mathbb{R} ^{N {\times} 1}$ is the SRP map, 
matrices $\mathbf{A}_k{\,=\,}[\mathbf{a}_1(k),{\cdots},\mathbf{a}_L(k)]{\,\in\,}\mathbb{C} ^{N {\times} L}$ and $\mathbf{B}_k=[\mathbf{b}_1(k),{\cdots},\mathbf{b}_L(k)]^{\top}{\in\,}\mathbb{C} ^{L {\times} Q}$ are formed by vectors $\mathbf{a}_\ell$ and $\mathbf{b}_\ell$, respectively,  $\mathbf{S}_{k{,}t}{\,=\,}[S_1(k{,}t),{\cdots},S_Q(k{,}t)]^\top{\in\,}\mathbb{C}^{Q {\times} 1}$ denotes the source weight matrix of $Q$ candidate positions, and $\boldsymbol{\mathcal{N}}_{k,t}{\in\,}\mathbb{R} ^{N {\times} 1}$ denotes the noise term.

Since sources are spare in numbers, we here assume $\mathbf{S}_{k,t}$ should have mostly zero entries. Under this assumption, we define equation \eqref{eq:newsrp2} as an underdetermined system where $Q{\,\geq\,}N$. This indicates that we represent SRP maps using a higher-resolution source weight map.
Hence, \eqref{eq:newsrp2} is a sparse representation of the SRP map for a single time frame and a single wave number. Such a sparse solution of $\mathbf{S}_{k,t}$ can be obtained by solving the following optimization problem:
\begin{equation}
\underset{\mathbf{S}_{k, t}}\min\frac{1}{2}\left\| \mathbf{z}_{k,t}-\Re(\mathbf{A}_k\mathbf{B}_k\mathbf{S}_{k,t})\right\|_2^2+\lambda\left\|\Re(\mathbf{S}_{k,t})\right\|_\mathfrak{p}^\mathfrak{p},
\label{eq:srpsbl1}
\end{equation}
where $\left\|\cdot\right\|_\mathfrak{p}$ denotes the $\ell_\mathfrak{p}$-norm as $0{\,<\,}\mathfrak{p}{\,\leq\,}1$.

To further enhance the robustness, we consider incorporating time-frequency diversity from SRP inputs. We here represent the SRP function as a multidimensional matrix $\mathbf{Z}{\,\in\,}\mathbb{R} ^{N {\times} T {\times} K}$ instead of the vector form $\mathbf{z}{\,\in\,}\mathbb{R} ^{N {\times} 1}$ in equation \eqref{eq:srpl10n}. We then express a group-sparse representation of the multidimensional matrix $\mathbf{Z}$ as
\begin{equation}
\mathbf{Z}=\Re(\mathbf{ABS})+ \boldsymbol{\mathcal{N}},
\label{eq:newsrptf}
\end{equation}
where $\mathbf{Z} \in\mathbb{R}^{N {\times} T {\times} K}$, 
$\mathbf{A} \in\mathbb{C}^{N {\times} L {\times} K}$, 
$\mathbf{B} \in \mathbb{C} ^{L {\times} Q {\times} K}$, \newline
$\mathbf{S} \in \mathbb{C}^{Q {\times} T {\times} K}$, and
$\boldsymbol{\mathcal{N}} \in \mathbb{R}^{N {\times} T {\times} K}$. Equation \eqref{eq:newsrptf} represents multidimensional matrix multiplication.

We solve the group-sparse representation \eqref{eq:newsrptf} by introducing a multidimensional mixed-norm penalty term \cite{mixednorm} as
\begin{equation}
\underset{\mathbf{S}}\min\frac{1}{2}\left\| \mathbf{Z}-\Re(\mathbf{A}\mathbf{B}\mathbf{S})\right\|_2^2+\lambda \mathcal{J}_{\mathfrak{p}_1,\mathfrak{p}_2,\mathfrak{p}_3}(\Re(\mathbf{S})),
\label{eq:srpsbl2}
\end{equation}
where $\mathfrak{p}_1, \mathfrak{p}_2, \mathfrak{p}_3$ denote the norms of spatial, time, and frequency domain, respectively, following a descending order where $0 {\,<\,} \mathfrak{p}_1 {\,\leq\,} \mathfrak{p}_2 {\,\leq\,} \mathfrak{p}_3 {\,\leq\,} 2$, and $\mathcal{J}_{\mathfrak{p}_1,\mathfrak{p}_2,\mathfrak{p}_3}(\Re(\mathbf{S}))$ denotes the penalty function of $\Re(\mathbf{S})$.

Here we assume source signals are non-sparse in time-frequency domain, hence we set $\mathfrak{p}_1{\,=\,}\mathfrak{p}, \mathfrak{p}_2{\,=\,}2, \mathfrak{p}_3{\,=\,}2$, which means the penalty term of \eqref{eq:srpsbl2} is a $\ell_{\mathfrak{p},2,2}$-norm function.

We next apply M-SBL to optimize \eqref{eq:srpsbl2} since it is suitable for solving the underdetermined problem without the prior knowledge of the source quantities $J$. 
The M-SBL method is based on two assumptions. First, the noise term $\boldsymbol{\mathcal{N}}$ is assumed to be the zero-mean Gaussian noise term with density $\mathcal{N}(\boldsymbol{\mathcal{N}} ; \mathbf{0}, \sigma^2 \mathbf{I})$. Second, the source weight $\mathbf{S}$ is assumed to be the zero-mean complex Gaussian with density  $\mathcal{C N}(\mathbf{S} ; \mathbf{0}, \mathbf{\Gamma})$, where $\mathbf{\Gamma}=\text{diag}(\boldsymbol{\gamma})$ is the diagonal matrix of hyperparameters $\boldsymbol{\gamma}=[\gamma_1,{\cdots},\gamma_Q]$.

In the M-SBL framework, the hyperparameters $\boldsymbol{\gamma}$ are assumed to be unknown and learned iteratively through maximizing the evidence to reach a sparse result. Here, the evidence model is also the zero-mean Gaussian with density $\mathcal{N}(\mathbf{Z} ; \mathbf{0}, \boldsymbol{\Sigma}_k)$, where $\boldsymbol{\Sigma}_k=\sigma_k^2\mathbf{I}+\mathbf{G}_{k}\mathbf{\Gamma}\mathbf{G}_{k}^H$. Maximizing $\hat{\boldsymbol{\gamma}}$ is facilitated iteratively using derivatives of the evidence. Further procedural details of M-SBL can be found in \cite{sbl_3}. Finally, the update equation is obtained as
\begin{equation}
    {\gamma}^{\text{new}}_q = \frac{\gamma_n^{\text {old}}}{T}\frac{{\textstyle\sum_{k=1}^{K}}\left\|\mathbf{Z}_k^H \boldsymbol{\Sigma}_k^{-1} \mathbf{D}_{k,q}\right\|_2^2}{{\textstyle\sum_{k=1}^{K}}\mathbf{D}_{k,q}^H\boldsymbol{\Sigma}_k^{-1}\mathbf{D}_{k,q}},
    \label{eq:SBL}
\end{equation}
where $\mathbf{D}_{k,q}$ denotes the $q^\text{th}$ column of the dictionary matrix $\mathbf{D}_{k} {\,=\,} \Re(\mathbf{A}_k\mathbf{B}_k)$. The candidate positions corresponding to the largest peaks of $\hat{\boldsymbol{\gamma}}$ will be the estimated source locations. It is worth noting that M-SBL terminates iterations when the error falls below the convergence threshold, thus removing the need to determine the number of sources. 

Compared to the current state-of-the-art sparsity-based modeling method for SRP maps \cite{srp_sparse_2}, the proposed method represents obtained SRP maps as a multidimentional matrix $\mathbf{Z}$ instead of a time-frequency-averaged vector $\mathbf{z}$ to improve localization by retaining time-frequency information. Additionally, we set two candidate grids for SRP mapping and sparse fitting with different resolutions for efficiency. We will compare these two methods and other localization methods through experiments in the next section.

\section{Experimental Validation}
\label{sec:typestyle}

\begin{figure}[t]
\centering
\subfigure[]{\includegraphics[width=0.4\columnwidth,clip,trim={2cm 12cm 2cm 1cm}]{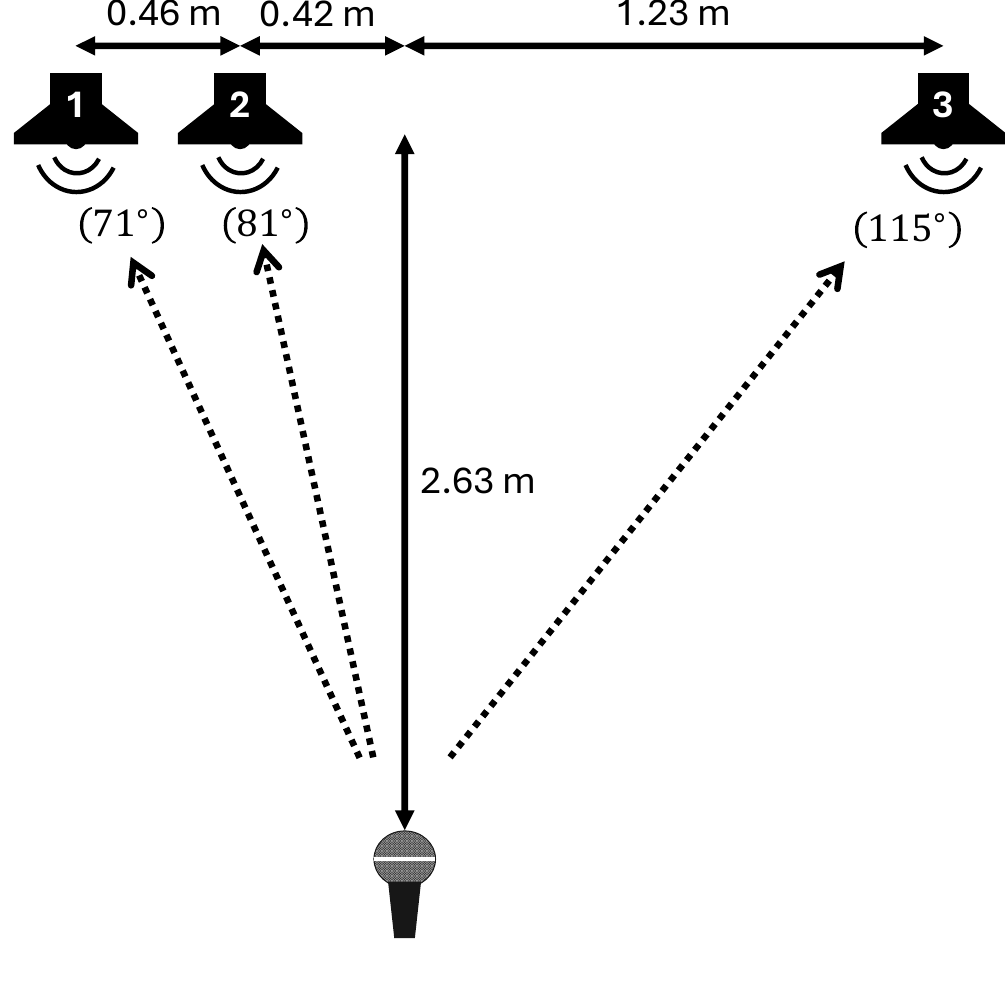}}
\subfigure[]{\includegraphics[width=0.5\columnwidth,clip,trim={0.1cm 13cm 1cm 1cm}]{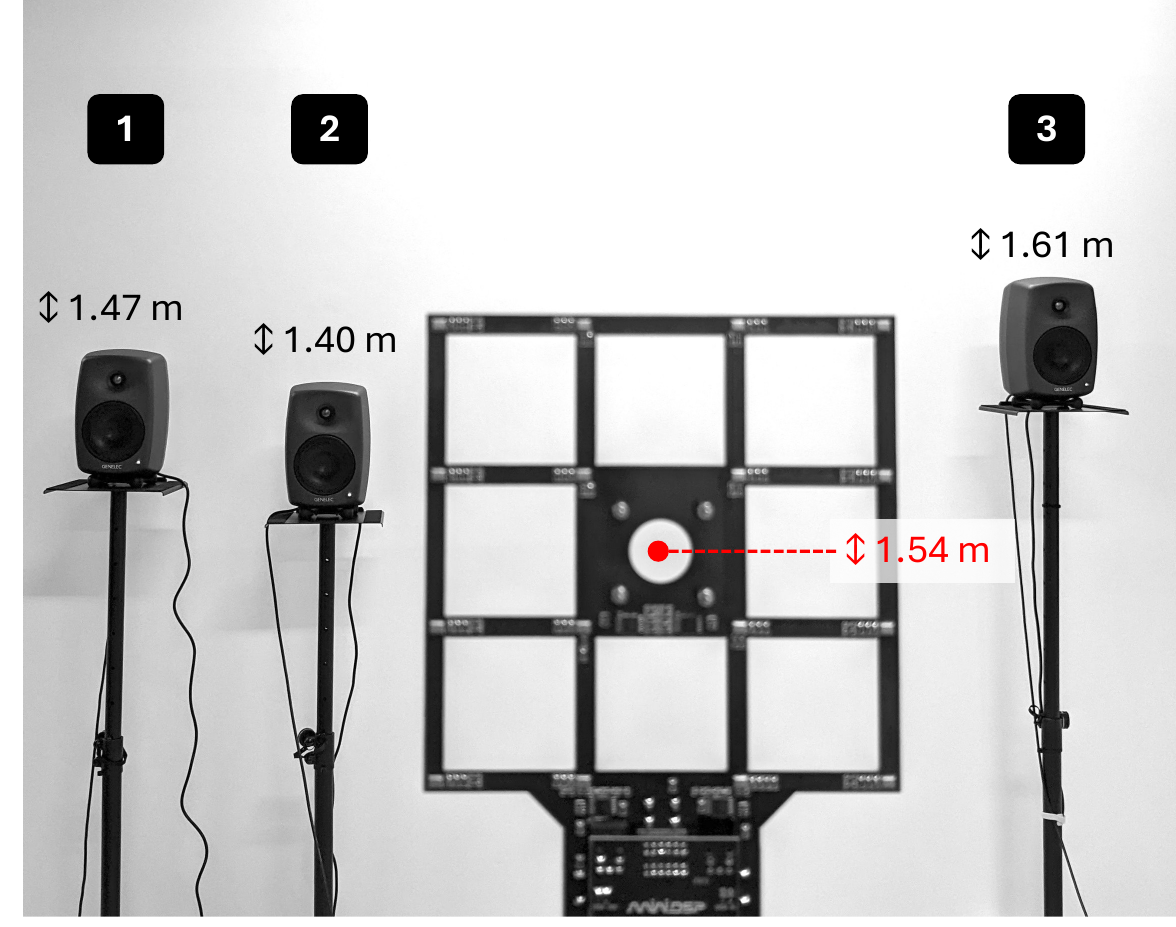}}
\caption{Illustration of experiment setup. Loudspeaker distance and azimuth positions are drawn in (a). Loudspeaker and microphone heights are pictured in (b).}
\vspace{-0.3cm}
\label{fig:placement}
\end{figure}

We conducted recordings in a large meeting room with a hard floor and ceiling at $3.3\text{ m}$ height, with a $T_{60}{\,\approx\,} 0.7\text{ s}$ ($T_{20}{\,=\,} 113\text{ ms}$ measured). We used a $M{\,=\,}16$-channel MEMS microphone planar array (MiniDSP UMA-16 v2) and $J{\,=\,}3$ target loudspeakers (Genelec 8030C) positioned in the room as detailed in Fig. \ref{fig:placement}. The loudspeaker's played speech signals from the MS-SNSD dataset \cite{ms-snsd}. The recording and localization used 48 kHz sampling frequency and 1024 STFT frame length with 50$\%$ overlap.
We note that the ground truth positions and DOAs of the speakers in Fig. \ref{fig:placement} were obtained by hand and are thus prone to human error.

We evaluate the DOA estimation performance of the proposed method through this experiment. 
We therefore use FF propagation for equations \eqref{eq: tdoa} and \eqref{eq:green}. The candidate DOAs are defined over a hemisphere since the planar array struggles in distinguishing front-back symmetry. We select $N{\,=\,}247$ candidate DOAs for the SRP map, with a $15^\circ$ resolution over elevation range $\theta{\,\in\,}[-90^\circ,90^\circ]$ and a $10^\circ$ resolution over azimuth range $\phi{\,\in\,}[0^\circ,180^\circ]$, $Q{\,=\,}8281$ candidate DOAs for the M-SBL, with a $2^\circ$ resolution for elevation $\theta$ and a $2^\circ$-resolution for azimuth $\phi$. The grids are with respect to the center of the microphone array. We set the convergence error of M-SBL as $10^{-3}$. We denote the proposed method as SRP-SBL in the results for brevity.

\begin{figure}[t]
\centering
\subfigure[]{\includegraphics[width=0.46\columnwidth,clip,trim={2cm 7cm 2cm 7cm}]{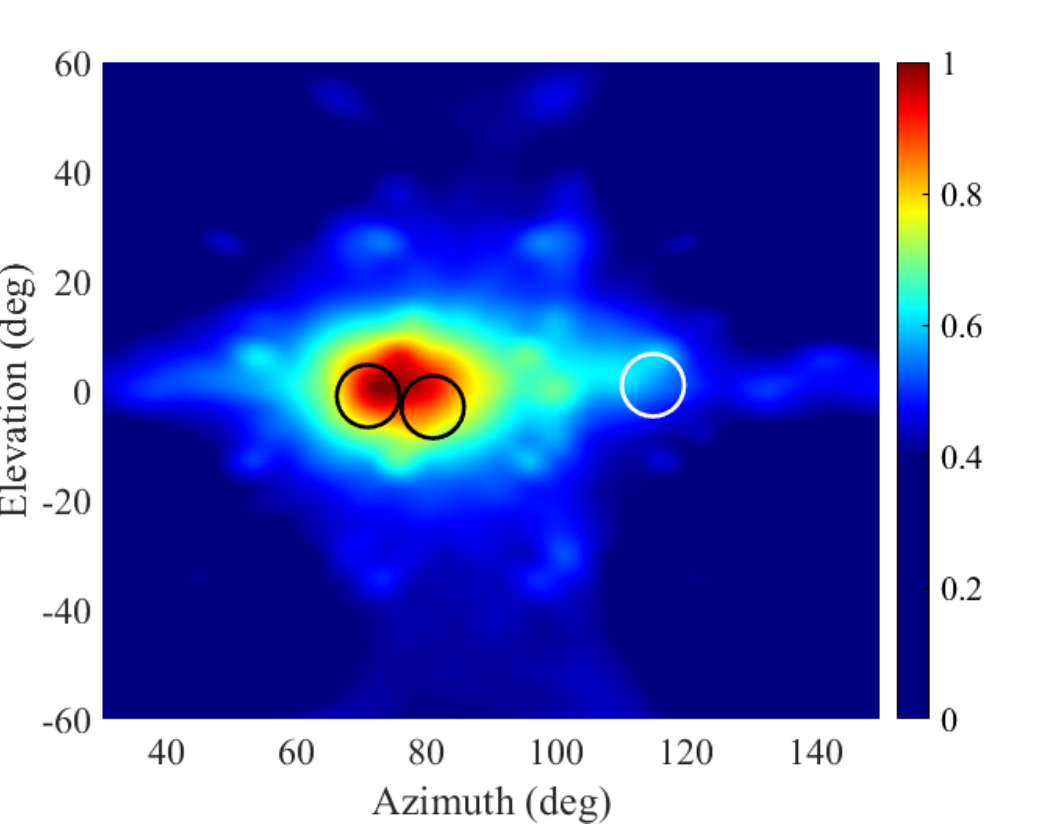}}
\subfigure[]{\includegraphics[width=0.46\columnwidth,clip,trim={2cm 7cm 2cm 7cm}]{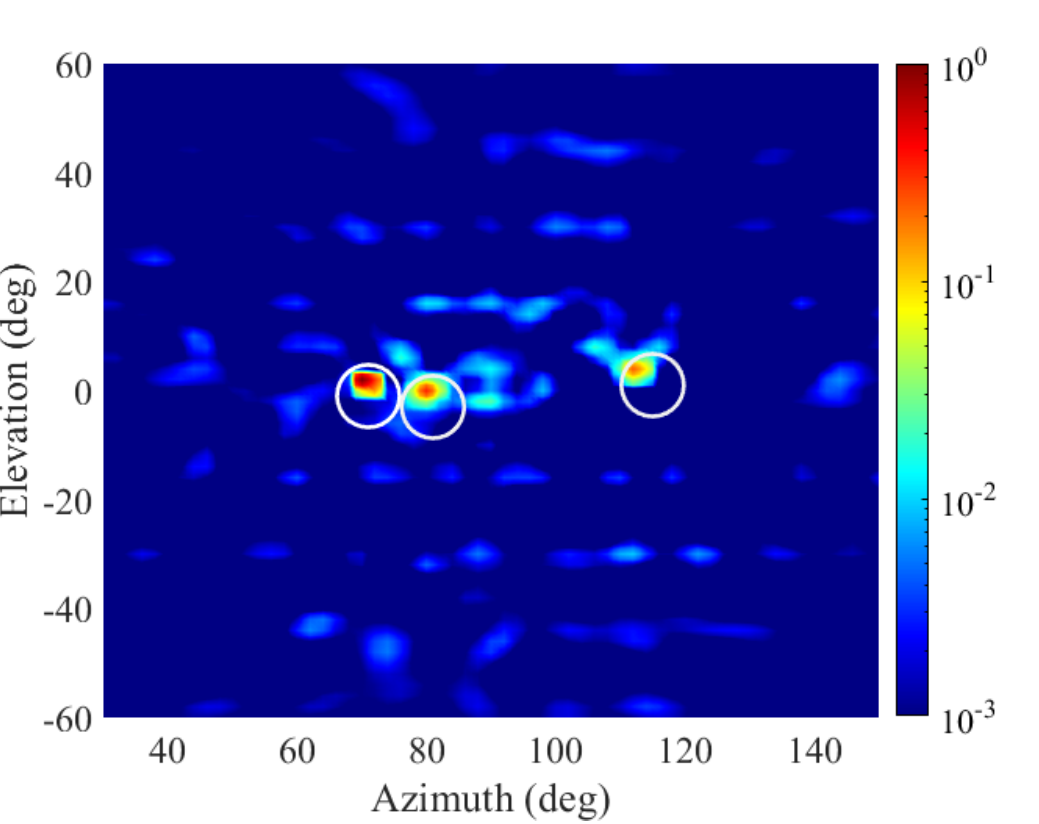}}
\caption{Normalized output maps of (a) SRP-PHAT and (b) SRP-SBL. Note that the color scale of (b) is logarithmic to better illustrate the localized peaks. The true DOAs are denoted by $(\circ)$.}
\vspace{-0.3cm}
\label{fig:map}
\end{figure}

We compare SRP-SBL to three methods: conventional SRP-PHAT, M-SBL, and SRP-sparsity (SRP-S)\cite{srp_sparse_2} which is the current state-of-the-art sparsity-based modeling method for SRP maps. The candidate DOAs for these methods are the same as SRP-SBL with $Q{\,=\,}8281$. 
It is important to distinguish that SBL utilizes microphone measurements $\mathbf{P}$ for sparsity optimization, as opposed to SRP-SBL which uses the SRP maps $\mathbf{Z}$.
Furthermore, the sparsity optimization method for SRP-S in \cite{srp_sparse_2} utilized the ADMM solver. However, due to the sensitivity of ADMM to adverse environments and the heuristic setting of the regularization parameter, we instead nominated to use Simultaneous OMP \cite{s-omp} as an alternative optimization method in our evaluation. 

In Fig. \ref{fig:map} we demonstrate the output maps of conventional SRP-PHAT and our proposed SRP-SBL. 
In Fig. \ref{fig:map}(a) we observe that the generated SRP-PHAT map exhibits an ambiguous peak at $(\theta{,} \phi)=(0^\circ{,}75^\circ)$, making it challenging to distinguish all three sources.
In contrast, the SRP-SBL map in Fig. \ref{fig:map}(b) clearly distinguishes the individual three sources and their relative locations. We note that the color-scale of this SRP-SBL map is logarithmic.
\begin{figure}[t]
    \centering
    \centerline{\includegraphics[width=7.5cm,clip,trim=0.5cm 7.8cm 0.5cm 8cm]{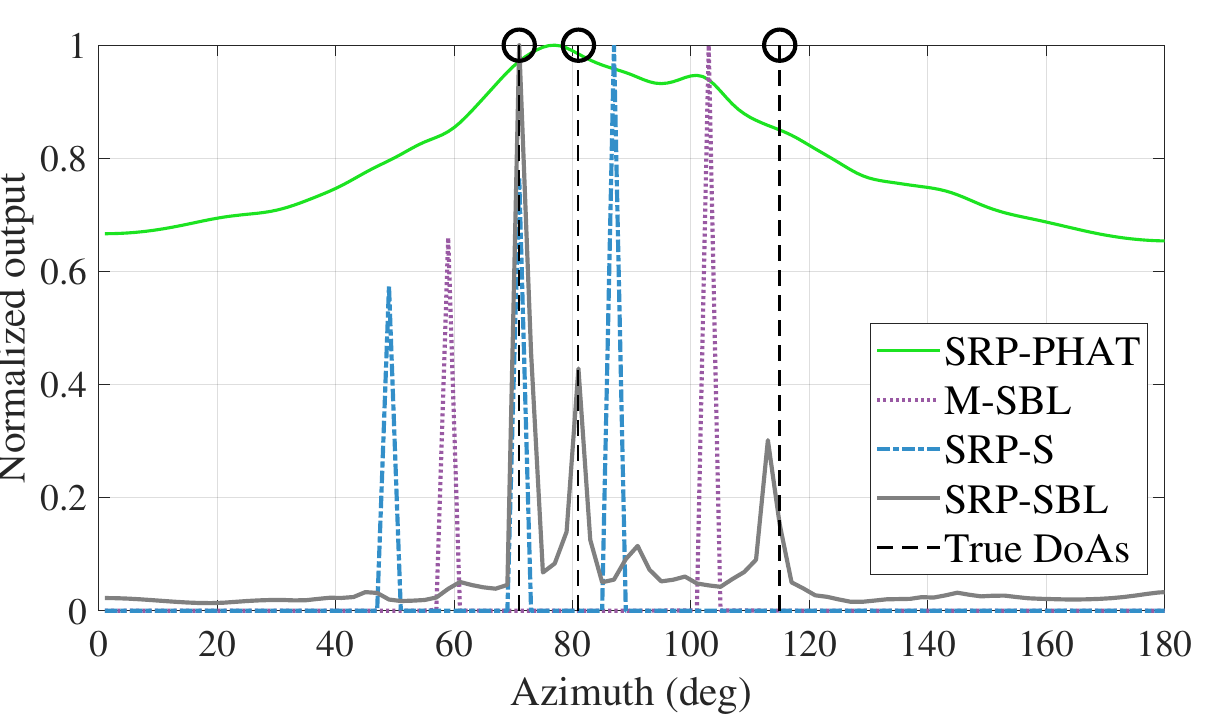}}
    \vspace{-0.3cm}
    \caption{Normalized output maps of the four methods and ground truth along the azimuth angles. The true DOAs are denoted by $(\circ)$.}
    \label{fig:2dmap}
    \centering
    \centerline{\includegraphics[width=7.5cm,clip,trim=0.5cm 7.8cm 1cm 8cm]{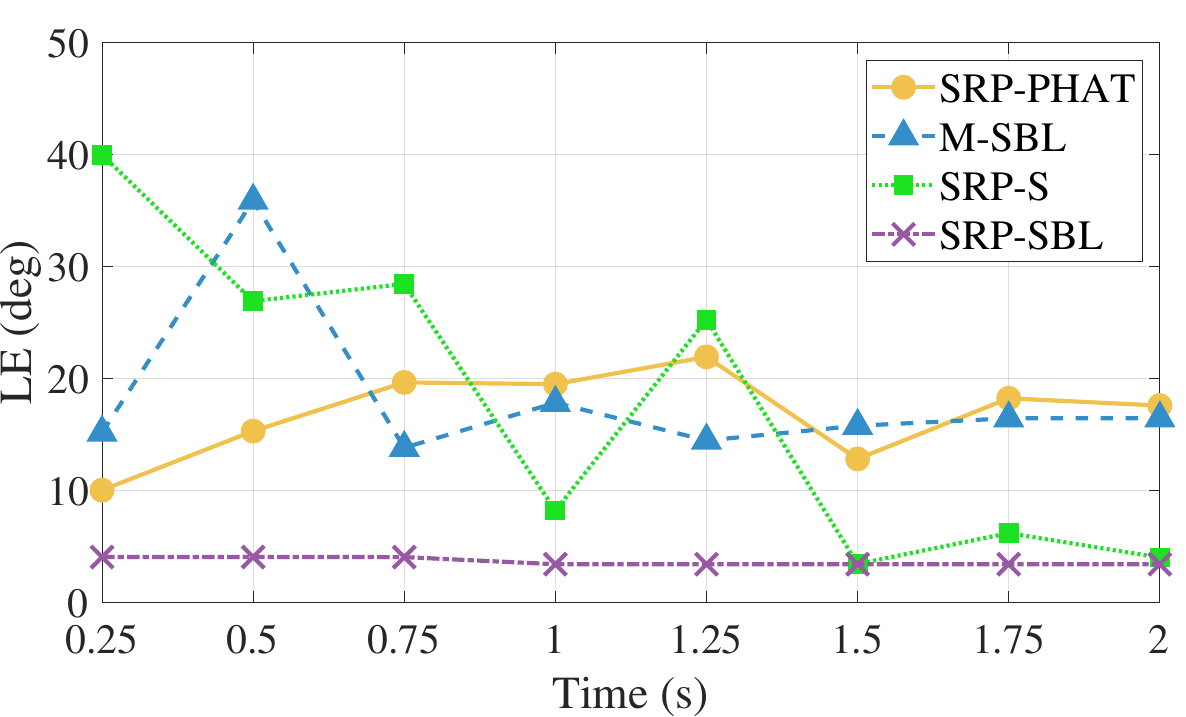}}
    \vspace{-0.3cm}
    \caption{LE for varying recording durations. }
    \vspace{-0.3cm}
    \label{fig:le}
\end{figure}

Fig. \ref{fig:2dmap} presents an azimuth slice of the Fig. \ref{fig:map} results along with the results of M-SBL and SRP-S.
We observe that M-SBL is unable to correctly localize the sources.
Whereas, SRP-S correctly localizes $\mathbf{y}_{1}$ and closely finds $\mathbf{y}_{2}$ but misses the third source.

We evaluate the impact of recording duration on localization accuracy in Fig. \ref{fig:le}.
In some scenarios where one source peak significantly outweighs others, leading to difficulties in estimating local maxima, we here impose a constraint that each estimated DOA should be spaced $3^\circ$ apart.
The localization error is given as the average angles between the estimated DOA $(\hat{\theta}_j{,}\hat{\phi}_j)$ of the $J$ highest peaks and the nearest true DOA $({\theta}_j,{\phi}_j)$ as
$
    \mathbf{LE}
    =
    \frac{1}{J} {\sum_{j\texttt{=}1}^{J}}
    \cos^{\shortminus 1}
    (\sin(\hat{\theta}_j)\sin({\theta}_j)\cos(\hat{\phi}_j{\shortminus}{\phi}_j)
    {+} 
    \cos(\hat{\theta}_j)\cos({\theta}_j))
$.
It can be observed that both SRP-PHAT and M-SBL yield significant $\mathbf{LE}$ for all recording duration.
The SRP-S method performs well for recording duration greater than $1\text{ s}$ but exhibits instability for duration ${\,\leq\,}1\text{ s}$. 
In contrast, the proposed SRP-SBL method is robust across a range of recording duration due to the way we maintain time and frequency diversity.


\section{Conclusion}
\label{sec:majhead}
In this paper, we propose a sparsity-based optimized SRP method for localizing multiple neighboring sources. The method utilizes a multidimensional SRP matrix as input and optimizes it through multidimensional SBL (M-SBL). Practical experiment results indicate that the proposed method outperforms conventional SRP, M-SBL, and the current state-of-the-art SRP sparsity-based method (SRP-S), maintaining stable localization performance in reverberant environments. In contrast, multiple closely spaced sources cause both SRP and M-SBL to consider all sources as a single point. While SRP-S enhances localization performance, it still loses its robustness in low-recording-duration scenarios.

\newpage \clearpage \newpage 
\balance
\bibliographystyle{IEEEbib}
\bibliography{refs}

\end{document}